\documentclass[%
 preprint,longbibliography,
 amsmath,amssymb,
 aps,prd
]{revtex4-1}

\usepackage[toc,page]{appendix}
\usepackage{longtable}
\usepackage{answers}
\usepackage{array}

\newcommand\fft[2]{\frac{#1}{#2}}
\newcommand\ft[2]{{\textstyle\frac{#1}{#2}}}
\newcommand\nn{\nonumber}
\renewcommand{\Re}{\operatorname{Re}}

\begin{document}
\preprint{LCTP-19-10}

\title{Gauged Supergravity from the Lunin-Maldacena background}

\author{James T. Liu}
\email{jimliu@umich.edu}
\affiliation{Leinweber Center for Theoretical Physics, Randall Laboratory of Physics, The University of Michigan, Ann Arbor, MI 48109-1040}

\author{Brian McPeak}
\email{bmcpeak@umich.edu}
\affiliation{Leinweber Center for Theoretical Physics, Randall Laboratory of Physics, The University of Michigan, Ann Arbor, MI 48109-1040}

\begin{abstract}
We show how to expand the Lunin-Maldacena solution to the full bosonic sector of gauged $\mathcal N=2$ supergravity.  In particular, we construct a consistent truncation of IIB supergravity on a $\beta$-deformed $\mathrm{AdS}_5\times S^5$ background retaining a dynamical metric and graviphoton in five dimensions.  We comment on the relation between this solution and similar constructions on Sasaki-Einstein manifolds, as well as the possibility of further consistent truncations which generalize the Lunin-Maldacena solution.
\end{abstract}

\maketitle


\section{Introduction}

It has long been known that $\mathcal{N} = 4$ super Yang-Mills (SYM) admits two exactly marginal deformations which break supersymmetry to $\mathcal{N}=1$ \cite{Leigh:1995ep}. These are the $\beta$ deformations and enter in the Lagrangian through $\mathrm{Tr} \, \left( \Phi_1 \Phi_2 \Phi_3 + \Phi_1 \Phi_3 \Phi_2 \right) $ and $ \mathrm{Tr} \, \left( \Phi_1^3 + \Phi_2^3 + \Phi_3^3 \right) $ terms in the superpotential. From an AdS/CFT point of view, these deformations should admit bulk duals which are solutions of type IIB supergravity on a correspondingly deformed $\mathrm{AdS}_5\times S^5$ background. However the problem of determining these solutions is highly non-trivial.

Marginal deformations in the field theory are generated by turning on massless scalars in the AdS dual that couple to operators with conformal dimension $\Delta=4$.  For $\mathcal N=4$ SYM, the dual is given by IIB supergravity on $\mathrm{AdS}_5\times S^5$, and its Kaluza-Klein (KK) spectrum has been extensively studied following the initial work of \cite{Gunaydin:1984fk,Kim:1985ez}.  At the lowest KK level, the 42 scalars of $\mathcal N=8$ gauged supergravity transform as $\mathbf{20}'+(\mathbf{10}+\overline{\mathbf{10}})+(\mathbf1+\mathbf1)$ of $\mathrm{SU}(4)_R$, with corresponding dimensions $2$, $3$ and $4$.  The latter complex scalar is identified as the IIB axi-dilaton, which is dual to the complexified gauge coupling, and generates an exactly marginal deformation of $\mathcal N=4$ SYM.

Additional massless scalars arise in the $\mathbf{45}+\overline{\mathbf{45}}$ representation at the first excited KK level, and in the $\mathbf{105}$ representation at the second excited level.  Of course, not all of these generate exactly marginal deformations, and the scalars dual to the $\beta$ deformation are fully contained in the $\mathbf{45}+\overline{\mathbf{45}}$ representation that arises as a KK excitation of the complex two-form potential of IIB supergravity.  The supergravity deformations resulting from these scalars were examined up to second order in \cite{Aharony:2002hx}, and subsequently a full non-linear construction of the $\beta$ deformation corresponding to $\mathrm{Tr} \, \left( \Phi_1 \Phi_2 \Phi_3 + \Phi_1 \Phi_3 \Phi_2 \right) $ was obtained by Lunin and Maldacena using an $SL(2,\mathbb R)$ solution generating method \cite{Lunin:2005jy}.  The second deformation is still unknown beyond second order.

The Lunin-Maldacena (LM) background preserves $\mathcal{N} = 2$ supersymmetry in five dimensions, so it is natural to expect that it can be extended to a full consistent truncation of IIB supergravity on a deformed $S^5$ resulting in five-dimensional $\mathcal{N} = 2$ gauged supergravity.  This would be in accord with the conjecture that any supersymmetric vacuum solution of KK form can be extended to a full non-linear KK reduction with the full set of corresponding supergravity fields active \cite{Duff:1985jd,Gauntlett:2007ma}.  We are particularly interested in the Lunin-Maldacena case because its starting point can be viewed as $\mathrm{AdS}_5\times S^5$ deformed by turning on a field in the `massive' KK tower.  Although the $\beta$ deformation is non-dynamical here, its presence nevertheless creates some tension between having non-trivial excitations in the KK tower and a consistent truncation that aims to remove such fields.

Of course, one can also take the point of view of gauged $\mathcal N=2$ supergravity in five dimensions.  In this case, we can consider the Lunin-Maldacena background as one preserving $\mathcal N=2$ supersymmetry and ignore the fact that massive $\mathcal N=8$ states are excited.  Then our aim is to consistently turn on an $\mathcal N=2$ graviphoton in this background.  At the linearized level, there is an obvious procedure for doing so by gauging the $U(1)_R$ isometry of the metric.  However, the non-linear reduction is not as straightforward.  Here, guided by the consistent truncation of IIB supergravity on a Sasaki-Einstein manifold \cite{Berenstein:2002ke,Buchel:2006gb}, we construct a full non-linear KK reduction to gauged $\mathcal N=2$ supergravity in the Lunin-Maldacena background.  While the Gauntlett-Varela conjecture \cite{Gauntlett:2007ma} has been verified for general AdS$_5$ solutions of M-theory \cite{Gauntlett:2004zh,Gauntlett:2006ai}, the present construction yields a non-trivial example starting directly from a IIB supergravity point of view.

\section{Reduction to $\mathcal N=2$ gauged supergravity}

We begin with the Lunin-Maldacena background, which, following the notation of \cite{Lunin:2005jy}, takes the form
\begin{align}
    ds^2&= G^{- 1/4} \left[ ds^2_{\mathrm{AdS}_5} + \sum_i( d\mu_i^2 + G \mu_i^2 d\phi_i^2)  + 9(\gamma^2+\sigma^2) G\mu_1^2 \mu_2^2 \mu_3^2 d \psi^2 \right], \nn\\
    e^{-\phi} &= G^{-1/2} H^{-1}, \qquad\chi =\gamma \sigma g_{0,E} H^{-1},\nn \\
    B_2 &= \ \gamma G w_2 - 12 \sigma w_1 \wedge d \psi , \qquad
    C_2 = \ -\sigma G w_2 - 12 \gamma w_1 \wedge d \psi ,\nn \\
    F_5 &=4(1+*)\omega_{\mathrm{AdS}_5}= 4(\omega_{\mathrm{AdS}_5}+Gdw_1\wedge d\phi_1\wedge d\phi_2\wedge d\phi_3),
\label{eq:LM}
\end{align}
where
\begin{align}
    G^{-1} & =  1 + (\gamma^2 + \sigma^2) g_{0,E}, \qquad H = 1 + \sigma^2 g_{0,E},\qquad g_{0,E} =\mu_1^2 \mu_2^2 + \mu_2^2 \mu_3^2 + \mu_3^2 \mu_1^2,\nn\\
    dw_1 &=\mu_1\mu_2\mu_3*_21, \qquad
    w_2 = \mu_1^2 \mu_2^2 d\phi_1\wedge d\phi_2 + \mu_2^2 \mu_3^2 d\phi_2\wedge d\phi_3 + \mu_3^2 \mu_1^2 d\phi_3\wedge d\phi_1.
\label{eq:GHg}
\end{align}
Here, we have written the five-sphere as a $T^3$ fibration over $S^2$, with $\{\phi_i\}$ as the torus coordinates and $\{\mu_i\}$ the `direction cosines' satisfying $\sum_i\mu_i^2=1$.  In addition, $\psi=(\phi_1+\phi_2+\phi_3)/3$ is the diagonal combination which defines the isometry direction dual to $U(1)_R$.  This solution is parametrized by two real constants, $\gamma$ and $\sigma$, which can be combined into a complex deformation parameter $\beta=\gamma-i\sigma$.  At linearized order, this deformation turns on the two-form potentials $B_2$ and $C_2$, which then backreact on the other fields in a manner that is consistent with \cite{Aharony:2002hx}.  Note that here we have chosen the initial IIB axi-dilaton to be $\tau=i$ prior to the $\beta$ deformation.

The one-form $w_1$ was introduced in \cite{Lunin:2005jy} as a potential, and is implicitly defined by its exterior derivative $dw_1$, where $*_21$ is the volume form on $S^2$.  In particular, for constant $\gamma$ and $\sigma$, only $dw_1$ shows up in the field strengths $H_3$ and $F_3$.  (See Appendix~\ref{app:IIB} for our IIB supergravity conventions.)  However, if the $\beta$ deformation were to be made spacetime dependent, then either $w_1$ would enter directly in the field strengths or some modification would be needed.  Although we do not pursue this approach here, we will nevertheless demonstrate below that including a dynamical graviphoton is sufficient to make a particular choice of $w_1$ physical.

\subsection{The reduction ansatz}

Although there is as yet no fully systematic treatment of consistent truncations, the starting point is clear as we can gain much insight from the linearized KK spectrum.  Since the deformed background (\ref{eq:LM}) preserves $\mathcal N=2$ supersymmetry, our aim is to obtain a truncation to the bosonic sector of $\mathcal N=2$ supergravity.  In particular, this involves the generalization of the AdS$_5$ background to an arbitrary five-dimensional space with metric $g_{\mu\nu}$ along with the addition of a graviphoton $A$ with field strength $F=dA$. 

To do this we will take advantage of the natural Sasaki-Einstein structure of $S^5$. Recall that Sasaki-Einstein manifolds are those which are both Sasaki and Einstein, and that a Riemannian manifold $\mathcal{S}$ is Sasaki if and only if its metric cone ($C = \mathbb{R}_{>0} \times \mathcal{S}, \  ds^2(C) = dr^2 + r^2 ds^2(S)$) is K{\"a}hler. The simplest example in five dimensions (and the one which is relevant for us) is the sphere, which has metric cone $\mathbb{C}^3 \backslash \{ 0 \}$. We will write the solution AdS$_5 \times S^5$ as a general Sasaki-Einstein compactification which retains the graviphoton, and then we will transform to the $\beta$-deformed theory.

Sasakian manifolds admit a Killing vector field known as the Reeb vector. When these orbits close, as is the case for the sphere, they define a foliation of $SE_5$. Then $SE_5$ may be written as a circle bundled over a four-dimensional K{\"a}hler base as:
\begin{equation}
    ds^2(SE_5) = ds^2(B) + \eta^2, 
\label{SE metric}
\end{equation}
with $d\eta=2J$ where $J$ is the K\"ahler form on the base.  In the case where $SE_5 = S^5$, the K{\"a}hler base is $CP^2$.

Since the graviphoton gauges the $U(1)_R$ isometry generated by $\partial/\partial\psi$, the metric ansatz is obtained by the replacement $d\phi_i\to d\phi_i+A$.  However, this is not yet complete, as the five-form field strength also gains graviphoton contributions in a Freund-Rubin setup.  In the absence of the $\beta$ deformation, a consistent Sasaki-Einstein truncation takes the form \cite{Berenstein:2002ke,Buchel:2006gb}
\begin{align}
    ds^2&=g_{\mu\nu}dx^\mu dx^\nu+ds^2(B)+(\eta+A)^2,\nn\\
    F_5&=(1+*)(4*_51-*_5F\wedge J)\nn\\
    &=4*_51+2J\wedge J\wedge(\eta+A)-*_5F\wedge J+F\wedge J\wedge(\eta+A),
\label{eq:SE}
\end{align}
where $*_5$ is the Hodge dual with respect to the five-dimensional metric $g_{\mu\nu}$. 

With (\ref{eq:SE}) as a starting point, we can turn on the Lunin-Maldacena deformation, which also brings the IIB axi-dilaton and two-form potentials into play.  The resulting ansatz takes the form
\begin{align}
    ds^2 &= G^{- 1/4} \left[g_{\mu\nu}dx^\mu dx^\nu + \sum_i(d\mu_i^2 + G\mu_i^2 (d\phi_i+A)^2)  + 9( \gamma^2+\sigma^2)G \mu_1^2 \mu_2^2 \mu_3^2 (d \psi + A)^2 \right],\nn \\ 
    \quad e^{-\phi}&= G^{-1/2} H^{-1}, \qquad \chi =  \gamma \sigma g_{0,E} H^{-1},\nn \\
    B_2  &= \gamma G w_2 - 12 \sigma w_1 \wedge \left( d \psi  + A \right),\qquad
    C_2  =  -\sigma G w_2 - 12 \gamma w_1\wedge \left( d \psi  + A \right),\nn \\
    F_5 &= 4*_51 + 4G d w_1\wedge(d\phi_1+A)\wedge(d\phi_2+A)\wedge (d\phi_3+A)\nn\\
    &\qquad- *_5F \wedge J + F  \wedge J \wedge(\eta+A) + 12 G \left( \gamma^2 + \sigma^2 \right) F \wedge w_1 \wedge w_2.
    \label{eq:GVansatz}
\end{align}
The scalar functions $G$, $H$ and $g_{0,E}$ are unchanged from (\ref{eq:GHg}), while $w_2$ now takes the form
\begin{equation}
    w_2=\mu_1^2\mu_2^2(d\phi_1+A)\wedge(d\phi_2+A)+\mu_2^2\mu_3^2(d\phi_2+A)\wedge(d\phi_3+A)+\mu_3^2\mu_1^2(d\phi_3+A)\wedge(d\phi_1+A).
\end{equation}
In addition, the forms pertaining to the Sasaki-Einstein structure can be expressed in terms of the $S^5$ quantities as
\begin{align}
    \eta+A &= \sum_i\mu_i^2(d\phi_i+A)=A + \sum_i\mu_i^2 d\phi_i,\nn\\
    2 J &= \sum_i d \mu_i^2 \wedge (d\phi_i+A)=\sum_id\mu_i^2\wedge d\phi_i.
\end{align}
Here we have made use of the constraint $\sum_i\mu_i^2=1$.

Note that the final term in the five-form ansatz in (\ref{eq:GVansatz}) is required by self-duality, as it is obtained by expanding out the ten-dimensional self-dual expression $F_5=(1+*)(4*_51-*_5F\wedge J)$ in the Lunin-Maldacena background.  It is interesting that the one-form $w_1$ appears directly, and not as a potential, in this term.  This is also the case for the three-form field strengths
\begin{align}
     H_3 &= \gamma \, G \, d w_2 - 12 \sigma\, dw_1\wedge(d\psi+A)+ 12 \sigma F \wedge w_1 - \gamma\left( \gamma^2 + \sigma^2 \right) \, G^2 \, d g_{0, E} \wedge w_2 ,\nn\\
    F_3 &= -\sigma \, H^{-1} \, d w_2 - 12 \gamma \, H^{-1}\,dw_1 \wedge(d\psi+A) + 12 \gamma  \, H^{-1}F \wedge w_1\nn \\
    & \kern4em + \sigma \left( \gamma^2 + \sigma^2 \right)  G \, H^{-1} \, d g_{0, E} \wedge w_2.
    \label{eq:3form}
\end{align}
As a result, turning on the graviphoton selects a preferred $w_1$ given as
\begin{equation}
    w_1 = -\frac{1}{12}\bigg[ (\mu_2^2 - \mu_3^2) \mu_1 d \mu_1 + (\mu_3^2 - \mu_1^2) \mu_2 d \mu_2 + (\mu_1^2 - \mu_2^2) \mu_3 d \mu_3 \bigg].
\end{equation}
It follows that
\begin{align}
    dw_1&=\fft13\left[\mu_1\mu_2d\mu_1\wedge d\mu_2+\mu_2\mu_3d\mu_2\wedge d\mu_3+\mu_3\mu_1d\mu_3\wedge d\mu_1\right]\nn\\
    &=\mu_1\mu_2\mu_3*_21,
\end{align}
where we have chosen an orientation such that
\begin{equation}
    *_2d\mu_i=\epsilon_{ijk}\mu_jd\mu_k,\qquad d\mu_i\wedge d\mu_j=\epsilon_{ijk}\mu_k*_21.
\end{equation}
From this point of view, $w_1$ is in fact physical, and can be expressed more compactly as
\begin{equation}
    w_1=\fft1{12}*_2d(\mu_1\mu_2\mu_3).
\end{equation}

\subsection{Verification of the ansatz}

We have verified that the above ansatz satisfies the IIB axi-dilaton and form field equations of motion.  Although we did not fully verify the IIB Einstein equation, we expect it to work as well.  The IIB equations of motion are satisfied provided the metric $g_{\mu\nu}$ and graviphoton $A_\mu$ obey the corresponding equations obtained from the bosonic Lagrangian of five-dimensional $\mathcal N=2$ gauged supergravity
\begin{equation}
    e^{-1}\mathcal L_5=R*_51+12*_51-\fft32F\wedge*_5F+F\wedge F\wedge A.
    \label{eq:5dim}
\end{equation}
The graviphoton kinetic term can be made canonical by the rescaling $A\to A/\sqrt3$.

In order to verify the ansatz, we had to compute the ten-dimensional Hodge dual of the field strengths.  This was done by splitting the ten-dimensional space into a warped product of five-dimensional spacetime, the $S^2$ base and the $T^3$ fiber
\begin{equation}
    ds^2=G^{-1/4}\left[g_{\mu\nu}dx^\mu dx^\nu+\sum_id\mu_i^2+G\left(\sum_ie_i^2+(\gamma^2+\sigma^2)\mu_1^2\mu_2^2\mu_3^2\Bigl(\sum_i\fft{e_i}{\mu_i}\Bigr)^2\right)\right],
\end{equation}
where $e_i=\mu_i(d\phi_i+A)$.  We use $*_5$, $*_2$ and $*_3$ to denote the Hodge duals within these three subspaces, respectively (without the overall $G^{-1/4}$ factor), and $*$ without subscript to denote the Hodge dual taken in the full ten-dimensional IIB metric (including $G^{-1/4}$).  In this case, we have the useful identities
\begin{align}
    *_31&=Ge_1\wedge e_2\wedge e_3,\nn\\
    *_3e_1 &= e_2 \wedge e_3 - G (\gamma^2 + \sigma^2) \mu_2 \mu_3 w_2,\nn \\
    *_3(e_1\wedge e_2)&=e_3+(\gamma^2+\sigma^2)\mu_1^2\mu_2^2\mu_3^{\vphantom2}\sum_i\fft{e_i}{\mu_i},\nn\\
    *_3(e_1\wedge e_2\wedge e_3)&=G^{-1},
\end{align}
along with cyclic permutations.  From these, we can obtain
\begin{equation}
    *_3\Bigl(\sum_i\fft{e_i}{\mu_i}\Bigr)=\fft{G}{\mu_1\mu_2\mu_3}w_2,\qquad*_3w_2=\mu_1\mu_2\mu_3G^{-1}\sum_i\fft{e_i}{\mu_i}.
\end{equation}

Verification of the form field equations of motion is straightforward although somewhat tedious.  Here we present some of the expressions that were useful in performing this check.  The IIB dilaton and RR scalar are naturally combined into the complex axi-dilaton
\begin{equation}
    \tau=\chi+ie^{-\phi}=(\gamma\sigma g_{0,E}+iG^{-1/2})H^{-1},
\end{equation}
with corresponding one-form field strength
\begin{equation}
    d\tau=-\ft12i(\sigma+i\gamma G^{1/2})^2H^{-2}G^{-1/2}dg_{0,E}.
\end{equation}
The three-form field strengths were given above in (\ref{eq:3form}), and can be combined into the complex three-form
\begin{align}
    G_3=F_3-ie^{-\phi}H_3&=(\sigma+i\gamma G^{1/2})H^{-1}\Bigl[-dw_2+4iG^{1/2}*_21\wedge*_3w_2-12iG^{-1/2}F\wedge w_1\nn\\
    &\kern9em+(\gamma^2+\sigma^2)Gdg_{0,E}\wedge w_2\Bigr],
\end{align}
with ten-dimensional Hodge dual
\begin{align}
    *G_3&=(\sigma+i\gamma G^{1/2})H^{-1}\Bigl[-G^{-1/2}*_{10}dw_2+4i*_51\wedge w_2+12i*_5F\wedge*_2w_1\wedge e_1\wedge e_2\wedge e_3\nn\\
    &\kern9em-(\gamma^2+\sigma^2)G^{1/2}*_51\wedge*_2dg_{0,E}\wedge*_3w_2\Bigr].
\end{align}
The axi-dilaton equation is then satisfied identically, while the three-form and five-form equations of motion are satisfied so long as the graviphoton satisfies the five-dimensional equation of motion $d*_5F=F\wedge F$ originating from (\ref{eq:5dim}).

\section{Discussion}

As demonstrated above, we have extended the Lunin-Maldacena solution into a full consistent truncation of IIB supergravity to the bosonic sector of pure five-dimensional $\mathcal N=2$ gauged supergravity.  It is of course interesting to ask if further consistent truncations generalizing the Lunin-Maldacena solution are possible.  Here there are several directions that this could take.  One possibility is to start from a consistent Sasaki-Einstein truncation \cite{Cassani:2010uw,Liu:2010sa,Gauntlett:2010vu,Skenderis:2010vz} restricted to the case of AdS$_5\times S^5$ and then turn on the Lunin-Maldacena deformation.  (Note that this can also be performed by a TsT transformation \cite{Frolov:2005dj} or a Yang-Baxter deformation \cite{Bakhmatov:2018apn}.) In addition to the pure supergravity fields, this would allow the retention of several additional multiplets in the consistent truncation.

As an example, we find that it is possible to retain a dynamical five-dimensional axi-dilaton $\tau_s=\tau_{1s}+i\tau_{2s}$ in the Lunin-Maldacena solution.  In fact, it can be shown that the solution of \cite{Lunin:2005jy} remains valid without modification, even for a dynamical $\tau_s$.  To demonstrate this, it is convenient to express the fields as
\begin{align}
    ds^2&= G^{- 1/4} \left[g_{\mu\nu}dx^\mu dx^\nu + \sum_i( d\mu_i^2 + G \mu_i^2 d\phi_i^2)  + 9\fft{|\beta|^2}{\tau_{2s}}G\mu_1^2 \mu_2^2 \mu_3^2 d \psi^2 \right], \nn\\
    e^{-\phi} &= \tau_{2s}G^{-1/2} H^{-1}, \qquad\chi =\tau_{1s}-\beta_1\beta_2 g_{0,E} H^{-1},\nn \\
    B_2 &= \fft{\beta_1}{\tau_{2s}}G w_2 - 12 \sigma w_1 \wedge d \psi , \qquad
    C_2 = \left(\beta_2+\fft{\tau_{1s}}{\tau_{2s}}\beta_1\right) G w_2 - 12 \gamma w_1 \wedge d \psi ,\nn \\
    F_5 &= 4(*_51+Gdw_1\wedge d\phi_1\wedge d\phi_2\wedge d\phi_3),
\end{align}
where
\begin{equation}
    G^{-1} = 1 + \fft{|\beta|^2}{\tau_{2s}} g_{0,E},\qquad
    H = 1 + \fft{\beta_2^2}{\tau_{2s}}g_{0,E},
\end{equation}
and we have introduced the shifted $\beta$-deformation parameter
\begin{equation}
    \beta=\beta_1+i\beta_2=\gamma-\tau_s\sigma.
    \label{eq:beta}
\end{equation}

A dynamical $\tau_s$ modifies the ten-dimensional one-form field strength
\begin{align}
    d\tau&=\fft{i}2(\beta_1+i\beta_2G^{-1/2})^2G^{1/2}H^{-2}dg_{0,E}\nn\\
    &\quad+\left(1+\fft{i\beta_1\beta_2}{\tau_{2s}}g_{0,E}G^{1/2}\right)H^{-1}\left(d\tau_{1s}+iG^{-1/2}H^{-1}d\tau_{2s}\right)+\fft{i}2\fft{\beta_2^2-\beta_1^2}{\tau_{2s}}g_{0,E}G^{1/2}H^{-1}d\tau_{2s},
\end{align}
as well as the complex three-form field strength
\begin{align}
    G_3&=\fft{\beta_1+i\beta_2G^{-1/2}}H\biggl[-iG^{1/2}dw_2-4G*_21\wedge*_3w_2+i\fft{|\beta^2|}{\tau_{2s}}G^{3/2}dg_{0,E}\wedge w_2\nn\\
    &\kern7.5em+\fft{G}{\tau_{2s}}\left((d\tau_{1s}-iG^{1/2}d\tau_{2s})+2i\fft{G^{1/2}H}{\beta_1+i\beta_2G^{-1/2}}(\beta_1d\tau_{2s}-\beta_2d\tau_{1s})\right)\wedge w_2\biggr].
\end{align}
The resulting equations of motion are then consistent with the five-dimensional Lagrangian
\begin{equation}
    e^{-1}\mathcal L_5=R*_51+12*_51-\fft1{2\tau_{2s}^2}d\tau_s\wedge*d\bar\tau_s.
\end{equation}
Although we have not included the graviphoton here, we expect that a full consistent truncation can be obtained that retains the complete set of fields of the generic squashed Sasaki-Einstein reduction.

Finally, in addition to $\tau_s$, we can also ask the question about whether it is possible to make the Lunin-Maldacena parameters $\gamma$ and $\sigma$ dynamical.  Like $\tau_s$, they are dual to exactly marginal deformations of the dual $\mathcal{N} = 4$ SYM theory.  However, there is a crucial difference in that they are part of the first excited KK level and moreover carry non-trivial dependence on the internal coordinates.  Stated differently, while it is always possible to obtain a consistent truncation by restricting to singlets under an internal symmetry group, in this case there is no such obvious subgroup that will retain $\gamma$ and $\sigma$ while removing the rest of the KK tower.

At the same time, however, the Lunin-Maldacena solution itself allows us to move continuously along the exactly marginal deformation parametrized by $\gamma$ and $\sigma$.  This raises the possibility that they may couple to higher states in the KK tower in a controlled manner.  After all, the truncation is consistent when these fields are set to constants, corresponding to turning on constant sources for the dual operators.  Additional motivation for a possible consistent truncation arises by noting that the shifted deformation parameter $\beta$ in (\ref{eq:beta}) can be spatially varying when $\tau_s$ is made dynamical.  This hints that an independent dynamical $\beta$ may be obtained using the dynamical $\tau_s$ solution as a starting point.  Nonetheless, we have found that a straightforward promotion of $\beta$ to an independently varying field does not lead to a consistent solution of the IIB equations of motion, so further study of the system will be required to see if such a truncation is possible.

\acknowledgments

This work was initiated following extensive discussions with U.\ Kol on supergravity duals of the $\beta$-deformations generalizing the Lunin-Maldacena background.  We also wish to thank C.N.\ Pope and E.\ Perlmutter for useful discussions on the possibility of a consistent truncation with a dynamical $\beta$-deformation parameter.  This work was supported in part by the U.S.~Department of Energy under grant DE-SC0007859.

\appendix

\section{Type IIB supergravity conventions}
\label{app:IIB}

The bosonic matter content of type IIB supergravity consist of the NSNS fields $g_{\mu\nu}$, $B_{\mu\nu}$ and $\phi$ and the RR fields $\chi\equiv C_0$, $C_2$ and $C_4$.  The field strengths obtained from these potentials are defined by:
\begin{align}
    F_1 &= d \chi, \qquad  \qquad H_3 = d B_2, \qquad  \qquad F_3 = dC_2 \, - \chi H_3,\nn \\
    F_5 &= d C_4 - \frac{1}{2} \, ( \, C_2 \wedge \, H_3 \, - B_2 \, \wedge \, d C_2 \, ),
\end{align}
where $F_5$ is self-dual, $F_5=*F_5$.  The Bianchi identities then follow:
\begin{align}
    d F_1 &= 0, \qquad \qquad d F_3 - H_3 \wedge F_1 = 0,\nn \\
    d H_3 &= 0, \qquad \qquad d F_5 - H_3 \wedge F_3 = 0
\end{align}

Type IIB supergravity does not admit a covariant action because of the self-duality of $F_5$.  However, we can write down the equations of motion.  The form field equations are
\begin{align}
    d( e^{2 \phi} * F_1) &=- e^{\phi} H_3 \wedge * F_3,\nn\\
    d*d\phi &= e^{2 \phi} F_1 \wedge * F_1 - \ft12 e^{- \phi} H_3 \wedge * H_3+\ft12 e^{\phi} F_3 \wedge * F_3,\nn \\
    d( e^{- \phi} * H_3) &=e^{\phi} F_1 \wedge * F_3 +F_3 \wedge F_5,\nn\\
    d( e^{\phi} * F_3) &=-H_3 \wedge F_5,
\end{align}
and the Einstein equation in Ricci form is
\begin{align}
    R_{\mu\nu}&=\fft12\partial_\mu\phi\partial_\nu\phi+\fft12e^{2\phi}\partial_\mu\chi\partial_\nu\chi+\fft14e^{-\phi}\left(H_{\mu\rho\sigma}H_\nu{}^{\rho\sigma}-\fft1{12}g_{\mu\nu}H_{\lambda\rho\sigma}H^{\lambda\rho\sigma}\right)\nn\\
    &\qquad+\fft14e^\phi\left(F_{\mu\rho\sigma}F_\nu{}^{\rho\sigma}-\fft1{12}g_{\mu\nu}F_{\lambda\rho\sigma}F^{\lambda\rho\sigma}\right)+\fft1{4\cdot4!}F_{\mu\lambda\rho\sigma\tau}F_\nu{}^{\lambda\rho\sigma\tau}.
\end{align}

The above equations can reexpressed in terms of the complex fields
\begin{equation}
    \tau=\chi+ie^{-\phi},\qquad G_3=F_3-ie^{-\phi}H_3.
\end{equation}
In particular, we have
\begin{align}
    \fft{d*d\tau}{\tau_2}+i\fft{d\tau\wedge*d\tau}{\tau_2^2}&=-\fft{i}{2\tau_2}G_3\wedge*G_3,\nn\\
    d*G_3&=-i\fft{d\tau}{\tau_2}\wedge*\Re G_3+iF_5\wedge G_3,\nn\\
    dF_5&=\fft{i}{2\tau_2}G_3\wedge \overline{G}_3,
\end{align}
along with
\begin{equation}
    R_{\mu\nu}=\fft1{2\tau_2^2}\partial_{(\mu}\tau\partial_{\nu)}\bar\tau+\fft14\left(G_{(\mu}{}^{\rho\sigma}\overline{G}_{\nu)\rho\sigma}-\fft1{12}g_{\mu\nu}G_{\lambda\rho\sigma}\overline{G}^{\lambda\rho\sigma}\right)+\fft1{4\cdot4!}F_{\mu\lambda\rho\sigma\tau}F_\nu{}^{\lambda\rho\sigma\tau}.
\end{equation}
Here the overlines indicate complex conjugation.

\bibliography{cite.bib}
\end{document}